\documentclass[%
 reprint,
 amsmath,amssymb,
 aps,
 prl,
 longbibliography,
 lengthcheck,%
]{revtex4-1}

\usepackage{graphicx}
\usepackage{dcolumn}
\usepackage{bm}
\usepackage{hyperref}

\begin{document}

\preprint{APS/123-QED}

\title{Thread-shaped intensity field after light propagation through the convective cell}

\author{Victor Alexeevich Kulikov${^1}$}
 \email{vkulik@mail.ru}
\affiliation{%
${^1}$ A.M.Obukhov Institute of Atmospheric Physics, Russian Academy of Sciences, Moscow 119017, Russia.}%

\author{Victor Ivanovich Shmalhausen${^2}$}
 \email{ vschm@optics.ru}
\affiliation{%
${^2}$ Chair of General Physics and Wave Processes, Department of Physics, M. V. Lomonosov Moscow State University, Moscow 119991, Russia.}%




\date{\today}

\begin{abstract}
A thread-shaped intensity field has been observed at 2 m distance when a laser beam passed through a water convection cell with characteristic Rayleigh number about $10^{8}$. Similar intensity fields were simulated by using phase screen method for various turbulent spectrum in the cell. We show that experimental result can be described by Tatarskii spectrum which take into account the inner scale.
\end{abstract}


\maketitle


\textit{Introduction.} The theory of turbulence has been introduced by Kolmogorov \cite{Kolmogorov 1941} and developed further by Obukhov\cite{Obukhov 1949} and Tatarskii \cite{Tatarskii 1961}. The theory was used to solve wide range of problems of light propagation through the turbulent mediums. Expressions that describe propagation of radiation in such mediums have been developed earlier using Rytov method in the approximation of weak fluctuations \cite{Rytov 1988}. Limitations of the theoretical scheme that postulates all fluctuations to be weak have been demonstrated in experiments by Gracheva and Gurvich \cite{Gracheva 1965}. Later on a lot of studies concerning propagation of radiation in a turbulent medium performed using approximate description of turbulence \cite{Lawrence 1970},\cite{Fante 1980}. The significance of the inner scale, which is responsible for the high frequency cutoff, and methods of its estimation were stated, for example in \cite{Hill 1992}. The effect of small spatial scales on the convection in Rayleigh-Benard cells is discussed in \cite{Lonse 2010}. Bissonete \cite{Bissonette 1977} passed a laser beam through a water cell, monitored the irradiance standard deviation and strength of turbulence $C_n^2$ and showed their compliance with the Kolmogorov's law prediction. The spectrum of refractive index diverge from Kolmogorov prediction in ocean turbulence because of salinity contribution \cite{Nikishov 2000},\cite{Korotkova 2012}.

Turbulence was studied also by methods of hydrodynamics in the cell \cite{Deardorf 1967}. Correct description of convective processes is based on the Navier-Stokes equations. Convection \cite{Siggia 1994},\cite{Brown 2007} is being studied both at theoretical \cite{Grossmann 2001} and experimental \cite{Ahlers 2012} levels. Spatial spectrum of the refractive index that depends directly on the temperature spectrum is being analyzed in typical optical studies. Hydrodynamic studies prefer to deal with velocity or temperature spectra \cite{Ashkenazi 1999}. Turbulence at high Rayleigh number values is active studied in different mediums such as air\cite{Li 2012}, helium\cite{Niemela 2000} and water\cite{Shang 2003}. The occurence of plume structures \cite{Gayen 2013}-\cite{Shishkina 2008} or, in other words, the clusterization \cite{Parodi 2004} of a 3D temperature field in convective medium at Rayleigh number values ($Ra=[10^{6}-10^{12}]$) has been reported in recent time.

\begin{figure}[h]
\includegraphics[width=8cm]{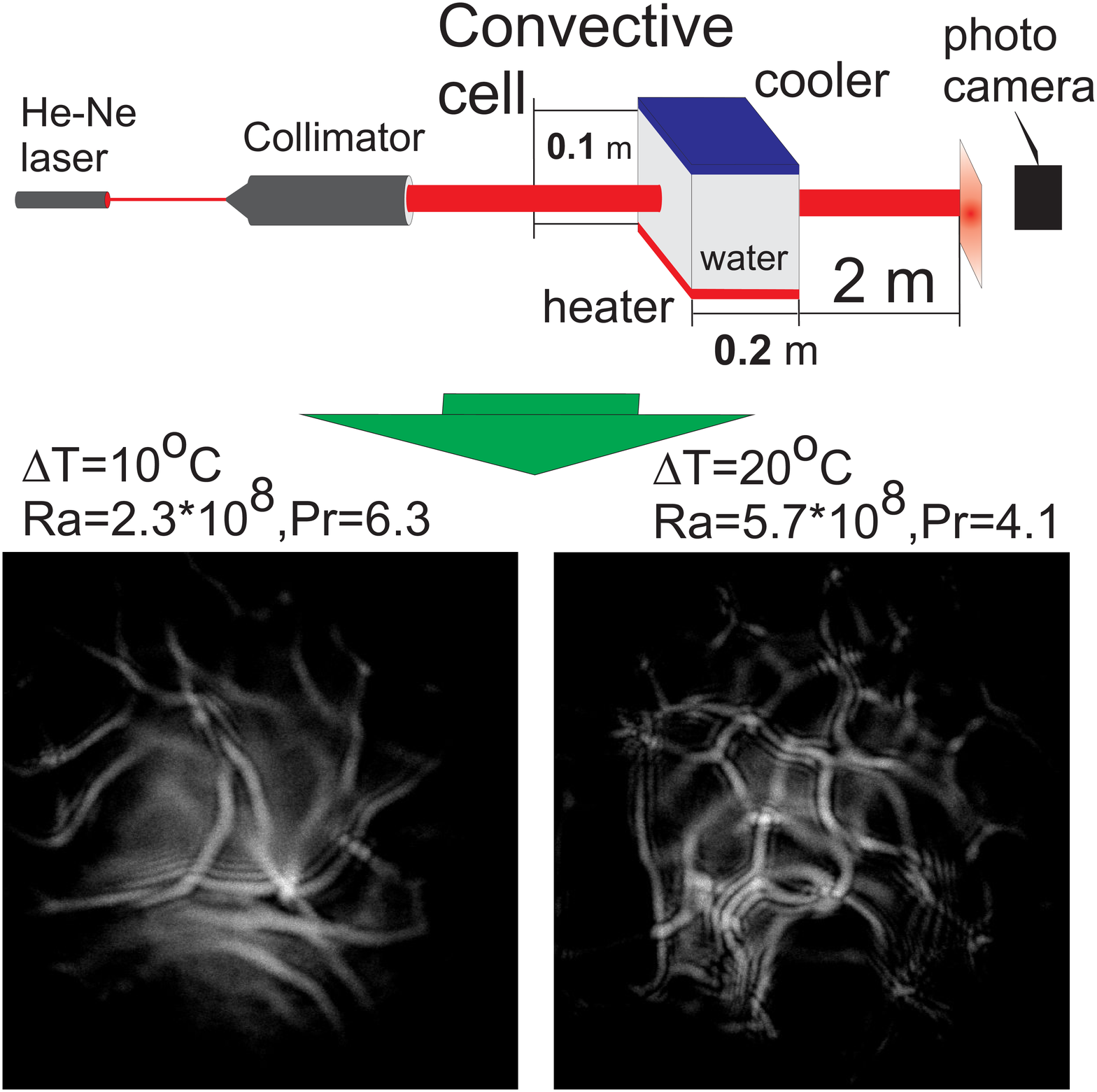}
\caption{ (color online) Experimental design and intensity field at the screen behind the cell.}
\end{figure}

We study the problem of description of light propagation through a convective water cell based on spectral description of turbulence. In this Latter we show that two-dimension Tatarskii spectrum is sufficient for simulation of the thread-shaped intensity fluctuation after light propagation through the turbulent water layer. Our result has also some academic significance because it demonstrates the capability of Kolmogorov theory to describe of the light propagation through the turbulence.

$\textit{Problem setup.}$ Experimental design is shown on fig.1. Convective cell is described in details in \cite{Kulikov 2012} and contains two horizontal plates: the lower heating plate and upper cooling plate which are held at constant temperatures. Turbulence has been induced with a vertical temperature gradient between plates. We passed a beam of collimated radiation which have 6 cm diameter through the cell filled with water.

Turbulence in the cell characterized the Rayleigh (Ra) number $Ra=(g\theta\Delta Tl^{3})/\nu\chi$ and the Prandtl (Pr) number  $Pr=\theta/\nu$ where $\textit{g}$ is the gravitational constant, $\textit{l}$ is a characteristic length ($\textit{l}$=10 cm), $\Delta T$ is temperature difference, $\nu$ is kinematic viscosity of the liquid, $\chi$ is thermal diffusivity, and $\theta$ is thermal expansion constant. The numbers of Rayleigh and Prandtl were equaled $Ra=2.3*10^{8}$ and Pr=6.3 for first and $Ra=5.7*10^{8}$ and Pr=4.1 for second regimes.
We used a 12-bit camera to get images of the beam on the screen at 2 m behind the cell. Relatively small distance from the cell leads to the conclusion that the intensity field we observe lies in Fresnel diffraction zone $(\lambda L)^{1/2} \ll D $, $\lambda=6,328*10^{-7}$ m, L=2 m, D=0,06 m).

Though convection cells have been used previously to study light propagation in turbulent media \cite{Gurvich 1977},\cite{Bissonette 1977} and to estimate turbulent characteristics \cite{Maccioni 1997},\cite{Kulikov 2012} plume structures have not been considered and discussed.

Both analytical and numeric approaches allow to investigate convection at Rayleigh numbers $Ra\sim10^{6}-10^{12}$ that contains plume structures are developed. They are based on Navier-Stokes equation solution and were described in \cite{Shiskina 2010}-\cite{Verzicco 2003}. Convection in rectangular cell at Rayleigh number $Ra=10^{8}$ contains plumes \cite{Gayen 2013},\cite{Shishkina 2008} which are presented for example on fig.3 in \cite{Shishkina 2008} or fig.1 in \cite{Gayen 2013}. We have not checked these structures in our experiment.

\textit{Numerical simulation.} Deviations from the Kolmogorov theory, which are observed in experiments, are under active study in recent time \cite{Dayton 1992},\cite{Zilberman 2008} and theoretical description of the problem is being developed \cite{Zilberman 2008},\cite{Toselli 2007}. We took the refractive index spectrum in its most general form which allows deviations from Kolmogorov theory \cite{Toselli 2007},
\begin{equation}\label{tp one} \Phi_n(k,\alpha)=A(\alpha) \tilde{C}_n^{2} exp[-(k^{2}/(k_m^{2}))] (k^{2}+k_0^{2} )^{(\alpha/2)}\end{equation}

 Conducting numerical simulations of the experiment, we used theory which supposed that both spectrum of temperature and spectrum of refraction are isotropic.

In this expression $3<\alpha<4$, k is the wave number, $k_0=2\pi/L_0$ and $L_0$ represents the outer scale, $k_m=c(\alpha)/l_0$, where $l_0$ is the inner scale, $c(\alpha)=[\Gamma((5-\alpha)/2)A(\alpha)2\pi/3]^{1/(\alpha-5)}$, $A(\alpha)=\Gamma(\alpha-1)cos(\alpha\pi/2)/4\pi^2$, $\Gamma(x)$ - gamma function, $\tilde{C}_n^{2}=\beta C_n^{2}$ is a generalized structure parameter with units $m^{-\gamma}$, $\gamma=\alpha-3$, $\beta$ is a function of $\alpha$ \cite{Toselli 2007}. If $\alpha=11/3$ ($A(11/3)=0.033$, $\tilde{C}_n^{2}=C_n^{2}$, $\beta=1$) then the expression (1) turns into Kolmogorov spectrum. If $l_0=0$ and $\alpha=11/3$ the expression (1) describes Tatarskii spectrum. To build a numeric model of the collimated beam propagation we used a phase screen method \cite{Martin 1990}-\cite{Martin 1988}, its precision has been discussed in detail in \cite{Martin 1988}. Taking the spectrum specified by (1) we generated random phase screens with given spectral properties.

The phase screen method supposes shrinking of a turbulent layer into infinitely thin phase screens in order to average irregularities over the thickness of the layer. In our experiment we had substantial phase distortions at output from the cell while the intensity almost did not vary. This allowed us to simulate turbulence as action of a single phase screen. Propagation of light to L=2.2 m has been performed with a split-step method.

\begin{figure}[h]
\includegraphics[width=8cm,angle=0]{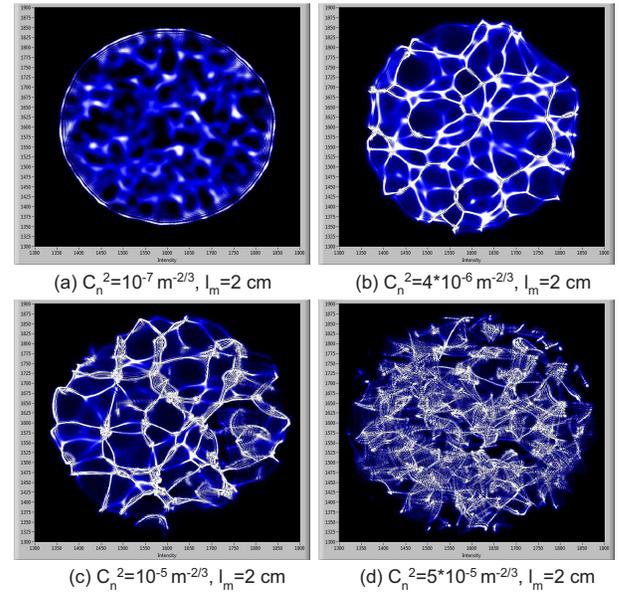}
\caption{ (color online) Numeric simulation: the intensity of a collimated beam at 2 meters behind the cell and its dependence on the turbulence strength $C_n^2$}
\end{figure}

By these means we obtained a set of statistically independent intensity fields, which similarly to the experimental pictures that have been shot at time intervals greater than the correlation time. We did not consider implementations, which are divided with shorter time intervals. Their simulation may possibly be performed with the spatial filtering method \cite{Koryabin 2006}.

We found that outer scale values $L_0>D$ do not affect the observed picture, where D is the beam diameter. So, all of the reported results have been obtained using the outer scale value $L_0=5D$. Diameter of the beam used was 6 cm, grid density – 8 points per millimeter. The field model consisted of $3200^{2}$ grid points.

\textit{Results}. Fig. 2 displays simulation results for particular implementations of the intensity field I at different $C_n^2$ while Fig. 3 shows results obtained at different $l_m$ values.
Speckle pattern is usually observed after a light beam propagation along atmospheric optical path. The intensity field, which is close to a speckle pattern, is shown at Fig. 2a. As the turbulence strength $C_n^2$ or the inner scale $l_m$ increases, the intensity field undergoes clusterization and formation of elongate thread-like structures can be observed. In relatively weak turbulence modes, which correspond to the onset of easily discerned thread-like pattern in the experiment (Fig.1) and to their occurrence in models (Fig.2b) and (Fig.3c), smooth high-intensive lines can be observed. As the turbulence strength $C_n^2$ increases, the structure of interference lines arises (Fig 2b, Fig. 3c). Note that modes described at Figs. 2a, 3a and 3d show no thread-like structures.

\begin{figure}[h!]
\includegraphics[width=8cm,angle=0]{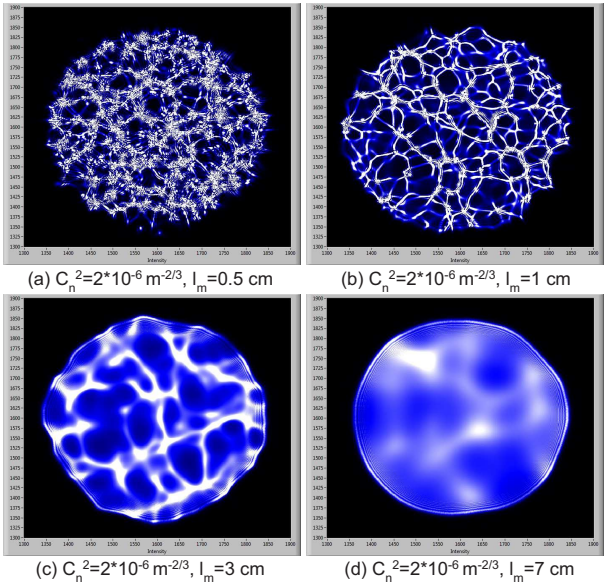}
\caption{ (color online) Numeric simulation: the intensity of a collimated beam at 2 meters behind the cell and its dependence on the inner scale $l_m$.}
\end{figure}

 Hence in the simulation for by using the Tatarskii spectrum the thread-like structure may appear at inner scale values from 1 cm to 5 cm and $C_n^2$ from $10^{-8}$ $m^{-2/3}$ to $10^{-5}$ $m^{-2/3}$. $C_n^2$ in water cells are known to be high, with typical observed magnitude $10^{-8}$ $m^{-2/3}$ [10]. Note, that simulation yields a thread-like intensity distribution regardless of the fact that phase additions field, which is calculated from the refraction index spectrum and gives rise to intensity fluctuations, is isotropic and does not contain such structures. When values of $C_n^{2}$ and $l_m$ reach $10^{-5}$ $m^{-2/3}$ and 5 cm, respectively, the thread-shaped distribution of the intensity field disappears due to strong interference.
Intensity fields similar to the one shown at Fig. 3a have been obtained in numeric simulation in \cite{Martin 1988} at relatively higher inner scale values.

To analyze statistical properties of obtained random fields we averaged the intensity over 500 implementations and considered a 350*350 points sized part of the beam, which corresponds to about $6*10^{7}$ points of the random field. Spatial spectrum of intensity field does not contain information about thread-shaped intensity distribution. Spectra of intensity fields, including thread-shaped, are totally isotropic in the range of parameters we consider, which means that the correlation function is also isotropic. This effect could be expected from random directions of threads that are visible at Fig. 2 and Fig. 3. Thus, it is possible to apply a single-dimension correlation (or structure) function to characterization and comparison of such kind of fields.
Fig. 4 presents plots of structure functions of the normalized intensity field $D(\mathbf{r}_1,\mathbf{r}_2)=\langle|\tilde{I}(\mathbf{r}_1) -\tilde{I}(\mathbf{r}_2)|^{2}\rangle$, where $\tilde{I}(\mathbf{r})=I(\mathbf{r})/\langle I\rangle$, against $C_n^2$ and $l_m$ values. Structure function raises as $C_n^2$ increases and falls with the increase of $l_m$. Increasing of the $C_n^2$ value leads to increased intensity fluctuation and the level of structure function saturation.

\begin{figure}[h]
\includegraphics[width=8cm,angle=0]{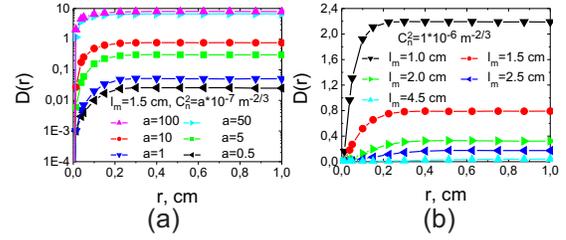}
\caption{ (color online) Plots of structural functions (simulation at $\alpha=11/3$) versus (a) $C_n^2$ at fixed $l_m=1.5$ cm (b) $l_m$ at fixed $C_n^2=1*10^{-6}$ $m^{-2/3}$. }
\end{figure}

The correlation length of the intensity field (the distance at which the structure function reaches saturation) depends on the inner scale magnitude. Increasing of the inner scale leads to the increase in the correlation length of the intensity field and slightly lowers fluctuation strength (Fig. 4b). Note that dependence of the fluctuation level on the inner scale is weak in the range involved $l_m=[1-5]$ cm. If $l_m$ value is fixed as at Fig. 4a the magnitude of fluctuations has obviously strong dependence of $C_n^2$ while the correlation length of the intensity field only slightly depends on $C_n^2$.

Our comparison of experimental and computed intensity fields is based on values of the normalized variance of intensity $\beta=\langle\tilde{I}^2\rangle/\langle \tilde{I}\rangle^2-1$ and the structure function $D(\mathbf{r}_1,\mathbf{r}_2)$. The saturation level of the structure function equals to doubled $\beta$. Our analysis of experimental data is restricted to the case of $Ra=2*10^{8}$.
Experimental data and numerical simulations based on Tatarskii spectrum are shown at Fig.5. Numerical dependencies presented at fig.5a were obtained at constant $l_m$ by variable $C_n^2$ values. Numerical dependencies shown at Fig.5b were obtained at constant $C_n^2$.

\begin{figure}[h!]
\includegraphics[width=8cm,angle=0]{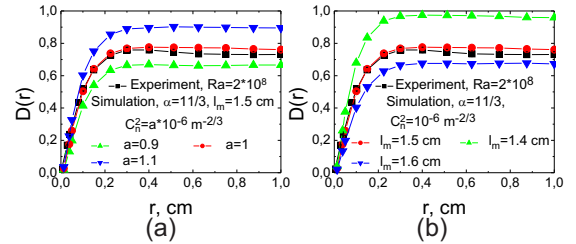}
\caption{ (color online) Estimates of experimental results at $\alpha=11/3$ (a) at fixed $l_m=1.5$ cm (b) at fixed $C_n^2=1*10^{-6}$ $m^{-2/3}$. }
\end{figure}

It's safe to say that our estimates of experimental parameters lie inside magnitude ranges [1.4-1.6] cm for $l_m$ and $[0.9-1.1]*10^{-6}$ $m^{-2/3}$ for $C_n^2$ (Table 1). At $l_m$ values $<1.4$ cm it's possible to fit a simulated random field with fluctuation variance close to the experimental value, the structural function in this case would, however, reach saturation faster and the correlation length of the intensity field would be less than that obtained in the experiment. At $l_m>1.6$ the situation is reversed.

The obtained estimation of the inner scale was in a good agreement with estimations obtained by using phase measurements in the convective cell \cite{Kulikov 2012}. Note that the inner scale and as well as the outer scale is needed to describe phase distortions \cite{Kulikov 2012}.

If the power low is changed (power coefficient $\alpha$ deviates from its normal value 11/3 in the range 3.5 to 3.8), the thread-shaped field effect may be simulated (Table 1) at other parameter ranges. If the power coefficient $\alpha$ raises the fluctuation level $\beta$ would decrease and we need higher $\tilde{C}_n^2$ values to account for our experimental data.

\begin{table}[h]
\caption{\label{qwerty}}
\begin{ruledtabular}
\begin{tabular}{cccc}
$\tilde{C}_n^2\cdot10^{-7}$, $m^{3-\alpha}$ & $l_m$, cm & $\beta$, simulation & $\beta$, experiment \\
\hline
\multicolumn{4}{c}{$\alpha=(11-0.5)/3$} \\
\hline
3-4 & 1.4-1.6 & 0.29-0.45 & 0.39$\pm$0.01 \\
\hline
\multicolumn{4}{c}{$\alpha=(11)/3$, Kolmogorov theory} \\
\hline
9-11 & 1.4-1.6 & 0.33-0.47 & 0.39$\pm$0.01 \\
\hline
\multicolumn{4}{c}{$\alpha=(11+0.5)/3$} \\
\hline
25-29 & 1.4-1.6 & 0.30-0.46 & 0.39$\pm$0.01 \\
\end{tabular}
\end{ruledtabular}
\end{table}

The length of the optical path has to be long enough for averaging when we used spectral method to describe light propagation through the convective turbulence. We showed that intensity scintillations produced by a turbulent water layer with the width equals 20 cm can be described in the framework of Kolmogorov theory in the terms of our experiment. The more long paths can be considered as made up of a set of the thin layers.

\textit{Conclusions.} The main idea of this letter is the possibility to build a description of the derived after the turbulent cell by means of numerical simulations based on Kolmogorov theory by taking the inner scale into account.

Authors give thanks to A.S. Gurvich for a fruitful discussion.

V.A. Kulikov thanks Russian Ministry of Education and Science for a grant $N$8613.

\end{document}